\documentstyle[12pt]{article}
\begin{document}
\begin{flushright}
FIAN/TD-4/95 \\
HU-TFT-95-13
\end{flushright}
\medskip
\begin{center}
{\bf{DILEPTON EMISSION FROM A RESONANCE GAS}}
\end{center}
\medskip
\begin{center}
{\bf{A.V.Leonidov$^{(1)}$
 and P.V.Ruuskanen$^{(2)}$}}
\end{center}
\begin{center}
{\it{$^{(1)}$ P.N.Lebedev Physics Institute, Moscow, Russia}}
\end{center}
\begin{center}
{\it{$^{(2)}$ Research Institute for Theoretical Physics, P.O.Box 9,
FIN-00014 University of Helsinki, Finland}}
\end{center}
\medskip
\begin{center}
{\it{Abstract}}
\end{center}
\baselineskip=18pt

We consider the dilepton emission from a dense and hot hadron gas near a
critical point. The hadron gas is described as a resonance one. The
dilepton spectrum generated by vector resonances is shown to be equivalent
to the one generated by quarks.

\newpage
\baselineskip=18pt

{\bf 1.} The dilepton production in high energy heavy ion collisions
has drawn considerable attention as a possible signal for the appearance
of a quark-gluon plasma [1], see also the reviews [2].
The main problem, however, is to
distinguish the signals coming from the "normal" hadronic phase and
from the quark-gluon plasma.
To be able to do so one needs reliable models for the dilepton production.
At extremely high temperatures the situation is relatively clear. Due
to the asymptotic freedom of QCD the dominant production mechanisms are
purely perturbative. At very low temperatures we also have a reliable
description of dilepton production in a rarefied hadron gas
(predominantly pions).
Much less can be said about the
"intermediate" temperatures range around the phase transition temperature
$T_c \sim 150$ MeV as given by the modern lattice data [3]
with a 15\% uncertainty.
We shall address the problem of estimating the dilepton emission from a
hadron gas at these temperatures. This is of special interest because
a hadron gas close to the transition temperature $T_c$ could be a dominant
feature of the final state produced at the present CERN heavy ion experiments
with lead beams.

{\bf{2.}} Near the phase transition point the hadron gas is
extremely dense and strongly interacting making the rigorous theoretical
analysis impossible. At the same time one could expect that in this case
its properties are very different from those of a gas of
free light hadrons. This also refers to the dilepton emission. Usually
one considers only a signal coming from a
rarefied hadron gas, so that only binary interactions are taken into
account (see, e.g., [4]).
Below we shall propose an estimate of a dilepton
signal coming from a hot dense hadron gas near a critical point.
The crucial idea of describing such a gas belongs to Hagedorn [5]
and for our purposes can be formulated as follows. The partition
function of a hot dense hadron gas can be rewritten as a sum of the
contributions
of an infinite number of heavy resonances. The solution of the corresponding
bootstrap equations [6] leads to a characteristic exponential mass spectrum:
\begin{equation}
\rho (m) = c {1 \over m^a} e^{m/T_0}
\end{equation}
%
% I changed crucial in the next sentence to essential to avoid repetition
% since crucial appears a coupple of sentences earlier.
%
The essential point is that the statistical bootstrap is being performed
only with respect to strong interactions, so one can consider the heavy
resonances saturating the partition function as a source of electromagnetic
and weak signals. In particular the heavy vector resonances provide a
natural source of high mass dilepton pairs. The mass spectrum of vector
mesons has the same shape as Eq.\ (1), with the normalization factor
$c_V$ determining its contribution to the total hadron spectrum.

Let us now consider the dilepton production by the vector resonances.
The dilepton width of a heavy vector meson is given by the standard
formula
\begin{equation}
\Gamma^{l^+ l^-}_V = {4 \alpha^2 \over 3} {m_V  \over g^2}
\end{equation}
where $\alpha$ is the fine structure constant, $m_V$ the mass of the vector
meson and $e m_V /g$ the coupling of the vector meson to the photon.
Leptons are assumed to be massless.
The spectrum of dilepton pairs produced by vector resonances
\begin{equation}
{d N^{l^+ l^-} \over dm d^4 x}=
\rho(m) \Gamma^{l^+ l^-} {3 m^2 T \over 2 \pi^2} K_2(m/T)
\end{equation}
is then a product of the mass spectrum, the thermal weight and the lepton
width of vector mesons.

{}From Eq.~(3) one could
expect,
that the exponential growth of a mass spectrum
will reduce the exponential damping coming from a thermal weight and therefore
increase the number of dileptons. Before making conclusions one should
however estimate the possible mass dependence of a vector meson-photon coupling
$g$. The simplest way to do this is to calculate the cross section of a
process $e^+ e^- \rightarrow V$.  Taking into account the mass
spectrum (1) and using the same photon coupling to vector mesons as in
Eq.~(2), we get
 \begin{equation}
\sigma (e^+ e^- \rightarrow V) ={(2 \pi)^3  \alpha^2 C_V \over  g^2}
 {1 \over Q^{1+a}}
e^{Q/T_0}
\end{equation}
where $Q$ is a CMS energy of $e^+ e^-$ collision which is equal to a mass
of produced vector meson.

Let us now assume an "absolute" vector dominance, namely that the cross
section of a vector meson production is equal to that of a total cross
section of $e^+ e^-$ annihilation into hadrons:
\begin{equation}
\sigma(e^+ e^- \rightarrow hadrons) = R {4 \pi \over 3} {\alpha ^2 \over Q^2}
\end{equation}
In the parton model at lowest order $R={5 \over 3}$ for the light quarks.
Equating the formulas for the cross section (4) and (5), we get for the
coupling constant
\begin{equation}
{1 \over g^2} ={2 \over 3} {R \over 2 \pi^2} {1 \over c_V} m^{a-1} e^{-m/T_0}
\end{equation}

We see that the mass dependence of the coupling turns out to be quite strong.
The leptonic decay width of a heavy vector meson goes down exponentially with
mass.  In the following we shall see that when combined with the bootstrap
mass spectrum, Eq.~(1), this does not mean that the dilepton radiation from
the hadron gas near the critical point is negligible, as it was recently
suggested in [7].  We should mention, however, that the status of the chiral
invariance in the bootstrap approach is not clear.

Substituting the expression (6) back into Eq.~(3) we finally get for the rate
\begin{equation}
{d N^{l^+ l^-} \over dm d^4x}={4R \alpha^2 \over 3(2 \pi)^{5/2}}
T^3 ({m \over T})^{3/2} e^{-m/T}
\end{equation}
which is exactly equal to the contribution at the same temperature from the
production of lepton pairs by quarks [8]!  We see that the exponentially
growing contribution from the spectrum is entirely compensated by the mass
dependence of the coupling constant $g$ which follows from the "absolute"
vector dominance assumption described earlier.

Let us mention, that the equality of rates from hadron gas and QGP reminds
the result of Kapusta {\it et al.} [9] who compared the rates for the
thermal photons.

{\bf{3.}} In this note we have shown that the consideration of a
dilepton emission by a hot and dense hadron gas near a critical point
leads to the same distribution in the invariant mass of produced
lepton pairs as for the dilepton production by quarks.

This result holds when averaged locally over resonances.
The low mass vector mesons can easily be resolved but in the mass range
above $\sim 1.5$ GeV we would expect that the resonance contribution is
smooth and cannot be distinguished from the quark contribution at the
transition temperature.
This regime should be relevant to the dilepton production in the
present experiments at CERN SPS collider with lead beams, where
the (hadronic) matter is expected to be produced with a temperature
of order of $T_c \sim 150$ Mev.
Let us note that the fact that we have obtained for the dilepton
production rate from a hadron gas the same answer as for
the production by quarks is entirely due to the "absolute" vector
dominance assumption in the $e^+e^-$ annihilation as described above.

We should like to emphasize that the above conclusion does not render the
dilepton signal useless as a probe of the dense stage of the collision.
First, if temperatures well above the transition temperatures are produced in
the heavy ion collisions, they will dominate the production of high mass
pairs.  Thus the slope and the intensity of high mass pairs is always related
directly to the properties of quark matter.  Second, even though the rate per
unit volume would be the same for hadron gas and plasma, the space-time volume
of the collision will depend on the equation of state of the matter [10].
This indicates that for a given multiplicity the total number of lepton pairs
will still depend on the dynamical properties of produced matter.

\begin{center}
{\it{Acknowledgements}}
\end{center}

A.L. is grateful to K.~Redlich and G.M.Zinovjev for useful discussions. His
work was supported by the Russian Fund for Fundamental Research, grant
93-02-3815.

\bigskip \bigskip

{\bf References}
\noindent

1. E.L.~Feinberg, {\it{Nuovo\ Cim.}} {\bf{A34}} (1976) 391;
  E.V.~Shuryak, {\it{Phys.\ Lett.}} {\bf{B78}} (1978) 150;
  G.~Domokos and J.I.~Goldman, {\it{Phys.\ Rev.}} {\bf{D23}} (1981) 203.

2. P.V.~Ruuskanen, {\it{Nucl.\ Phys}}\ {\bf{A544}} (1992) 169c;
   J.I.~Kapusta, {\it{Nucl.\ Phys.}}\ {\bf{A566}} (1994) 45c.

3. B.~Petersson, {\it{Nucl.~Phys.~B~[Proc.~Suppl.]}}\
{\bf {30}} (1993) 66; F.~Karsch and E.Laermann, {\it{Rep.~Progr.~Phys.}}\
{\bf{56}} (1993) 1347.

4. C.~Gale and  P.~Lichard, {\it{Phys.\ Rev.}}~{\bf{D49}} (1994) 3338.

5. R.~Hagedorn, {\it{Nuovo.\ Cim.\ Suppl.}}\ {\bf{3}} (1965) 117.

6. S.~Frautschi, {\it{Phys.\ Rev.}}\ {\bf{D3}} (1971) 2821.

7. D.~Kharzeev and H.~Satz, "Chiral Dynamics, Deconfinement and
Thermal Dileptons", preprint CERN-TH.7399/94.

8. K.Kajantie, J.Kapusta, L.McLerran and A.Mekjian,
{\it{Phys.Rev.}} {\bf{D34}} (1988) 2746.

9. J.~Kapusta, P.~Lichard and D.~Seibert, {\it{Phys.~Rev}}\
{\bf{D44}} (1991) 2744.

10. M.~Kataja, P.V.~Ruuskanen, H.~von Gersdorff and L.~McLerran, {\it
Phys.\ Rev.}\ {\bf D34} (1986) 2755.

\end{document}